# Electronic and magnetic properties of multiferroic ScFeO$_3$

S W Lovesey[1,2], and D D Khalyavin[1]

[1]ISIS Facility, STFC Oxfordshire OX11 0QX, UK

[2]Diamond Light Source Ltd, Oxfordshire OX11 0DE, UK

**Abstract**

Experimental techniques using Bragg diffraction may elucidate electronic and magnetic properties of ferric ions in multiferroic ScFeO$_3$ that are puzzling. A collinear magnetic motif of dipole moments is not expected for s-state ions with an acentric parent (paramagnetic) structure, because single-site anisotropy is negligible and antisymmetric exchange interactions promote an inhomogeneous magnetic motif. However, current indications are that ScFeO$_3$ supports G-type antiferromagnetism, with an unknown moment direction, which implies that ferric ions (Fe$^{3+}$) are not in the high-spin $^6$S state. It is argued that significant insight on these open questions can be achieved by confronting experimental data with calculated structure factors for resonant x-ray and magnetic neutron Bragg spots, and we report results for G-type antiferromagnetism allowed by the corundum-like parent R3c structure. Calculated structure factors include Dirac multipoles that are time-odd and parity-odd, e.g., dipoles that are often called anapoles or toroidal moments.

## I. INTRODUCTION

Bragg diffraction experiments on ScFeO$_3$ can address important questions about its electronic and magnetic properties that remain open despite previous thorough investigations [1, 2]. The unknown direction of the Fe magnetic moment in the basal plane of the corundum-like structure can be derived from resonant x-ray diffraction following a proposal by Chapon and Lovesey [3] that was subsequently proved in experiments on an iridate [4]. Similar experiments will be an acid test of a high-spin state for the ferric ion (3d$^5$), in addition to finding the moment direction. The reported collinear antiferromagnetism is unexpected [1], for one thing, because the single-ion anisotropy ought to be too small to gain favour over antisymmetric exchange interactions and a concomitant inhomogeneous magnetic motif. A magnetoelectric material supports Dirac multipoles that exist because of p-d hybridization, which is predicted to be very strong in ScFeO$_3$ [2]. Dirac multipoles, the best known of which is an anapole (toroidal dipole), fortunately deflect both x-rays and neutrons.

ScFeO$_3$ has been synthesized under high pressure and high temperature conditions [1]. The phase is described by the polar R3c crystal structure (Fig. 1a), and ferroelectric polarization is present at room temperature. Weak ferromagnetism with an ordering temperature $T_N \approx 545$ K is attributed to a canted G-type antiferromagnetic motif of Fe$^{3+}$ magnetic dipoles, two examples of which are depicted in Figs. 1b and 1c [1, 2]. In respect of the magnetic motif ScFeO$_3$ is apparently very different from the well-known multiferroic BiFeO$_3$, which has a high magnetic ordering temperature and large ferroelectric polarization [5, 6].

Actually, ScFeO$_3$ shares the same parent symmetry R3c1′ with multiferroic BiFeO$_3$. In the latter case, the polar nature of the crystal structure results in a long-period spin cycloid due to antisymmetric exchange interaction [5, 6]. This interaction competes with the single-ion anisotropy term, and in the symmetry-based phenomenological theory it is represented by a Lifshitz invariant [9]. The presence of this invariant, in the free-energy decomposition, is also allowed in the case of ScFeO$_3$, favouring the inhomogeneous magnetic state. The experimentally found homogenous G-type antiferromagnetic structure [1] signifies a predominant role of the single-ion anisotropy. This indicates that the actual electronic configuration of Fe$^{3+}$ in ScFeO$_3$ can be essentially different from the high-spin isotropic s-state. Thus, an experimental investigation of the electronic states of iron in ScFeO$_3$ can be an indispensable step toward a meaningful comparative analysis of the two high-temperature multiferroics BiFeO$_3$ and ScFeO$_3$.

In spite of the fact that the polar structure of ScFeO$_3$ is a high-pressure metastable phase and obtaining high-quality single crystals is very challenging, it might appear that the structure can be grown as epitaxial film using rather conventional methods [7].

Magnetic structures compatible with the polar R3c chemical structure and a G-type antiferromagnetic motif are discussed in the following section, with some details assigned to an Appendix. The two monoclinic candidates under consideration, Cc and Cc′, allow the magnetic phase-transition to be continuous [1]. Magnetic dipole moments for the structures are examined theoretically in Section III, which helps set the scene for discussions of resonant x-ray and magnetic neutron diffraction in the following two sections.

## II. MAGNETIC STRUCTURE AND SPACE GROUPS

Iron ions occupy sites 6a (0, 0, z = 0) in the acentric parent (paramagnetic) structure R3c1′ that is depicted in Fig. 1a [1]. There are three magnetic space-groups consistent with a G-type antiferromagnetic motif depending on the orientation of dipole moments, which is currently unknown. The triad axis of rotation symmetry at sites 6a in the hexagonal parent is removed in magnetic structures that use monoclinic cells. Magnetic space-groups are specified in terms of the Belov-Neronova-Smirnova notation [8].

The magnetic space-groups are as follows. Cc (#9.37 magnetic crystal-class m) in-plane Fe magnetic dipole moments are perpendicular to the glide plane: Cc′ (#9.39 magnetic crystal-class m′) magnetic moments are parallel to the glide plane: magnetic moments are in a general direction within the plane, and the magnetic space-group is P1 (#1.1). The experimental evidence of a classical second-order magnetic transition $T_N \approx 545$ K [1] is best described by one of the two monoclinic space-groups. This is because Cc and Cc′ are maximal isotropy subgroups that allow the transition to be continuous, whereas the triclinic subgroup, P1, is not a maximal one and requires a first-order transition.

In subsequent work we consider motifs described by Cc and Cc′, with properties of Fe ions referred to orthonormal principal axes (ξ, η, ζ) that are defined in an Appendix, which also contains information about P1 (#1.1). For the moment, it is sufficient to note that the η-axis of the monoclinic cell is parallel to a glide plane, and the ζ-axis subtends an angle (β − π/2) ≈ 32.857 ° with the hexagonal c-axis (β is the obtuse angle of the monoclinic cell). Monoclinic unit cells are depicted in Fig. 1b (Cc′) and Fig. 1c (Cc). Note that the same magnetic space groups are also expected for multiferroic BiFeO$_3$, where a G-type magnetic structure can be stabilized by an external magnetic field [9] or by chemical doping [10].

Regarding bulk properties, the two magnetic space groups have identical properties for non-magnetic phenomena and ferric ions can be displaced in ξ-ζ plane. We use **n** for the electric dipole operator and angular brackets ⟨ ... ⟩ to denote the time average, or expectation value, of the enclosed operator. Using this notation, ⟨**n**$_\xi$⟩ and ⟨**n**$_\zeta$⟩ may be different from zero. Available calculations say that displacements of the cations are minor compared to displacements of Sc and O ions that largely determine the spontaneous polarization along the hexagonal c-axis [2].

A ferromagnetic moment formed with μ$_\eta$ = ⟨(L + 2S)$_\eta$⟩ is allowed in Cc, whereas μ$_\xi$ and μ$_\zeta$ are allowed in Cc′. Anapoles (Dirac dipoles) are derived from magnetoelectric operators **S x n** and **Ω** = (**L x n** − **n x L**). A ferro-type order of anapoles using ⟨(**S x n**)$_\eta$⟩ and ⟨**Ω**$_\eta$⟩ is allowed in Cc′, while participating anapoles are in the ξ-ζ plane for Cc.

### III. MAGNETIC MOMENTS

We consider the magnetic dipole **μ** = ⟨(**L** + 2**S**)⟩ by way of an introduction to more subtle magnetic phenomena. The observed antiferromagnetic motif is consistent with Fe dipoles in the basal plane, perpendicular to the hexagonal c-axis, and μ ≈ 3.71 μ$_B$ at 5 K [1]. A suitable wavefunction for the ground-state of a ferric ion with such a dipole moment follows from the construction used by Chapon and Lovesey [3] in their study of an iridate compound, so we can now be brief.

With an odd number of electrons the ground-state can be taken to be a linear combination of Kramers' states,

$$|g\rangle = (1/\sqrt{2})\,[|u\rangle + f\,(\theta|u\rangle)], \qquad (1)$$

where θ is the time-reversal operator, f is a phase factor with $|f|^2 = 1$, and $|u\rangle$ is composed of states with half-integer angular momenta, J. Assuming a high-spin ferric ion, $|u\rangle$ is derived from states labelled by projections −J ≤ M ≤ J. For $|u\rangle = [\alpha|M\rangle + \delta|M'\rangle]$ and $|M − M'| > 1$ it follows that ⟨μ$_c$⟩ ≡ ⟨g|μ$_c$|g⟩ = 0, as required. In the present case, J = 5/2 and the observed saturation magnetic moment can be reproduced by M + M′ = 1 and M′ = M + 2, and real coefficients α, δ. Specifically,

$$\langle\mu_\eta\rangle = f''\,[3\alpha^2 − 4\sqrt{2}\,\alpha\delta], \qquad (2)$$

with $\alpha^2 + \delta^2 = 1$ and f″ = Im.f. This result and f′ = Re.f = 0 is compatible with Cc′. On the other hand,

$$\langle\mu_\zeta\rangle = −\cos(\beta)\,f'\,[3\alpha^2 + 4\sqrt{2}\,\alpha\delta], \qquad (3)$$

and f″ = 0 is compatible with Cc.

Foregoing results are derived with the aid of a standard operator equivalent **μ** = g**J**, where g is the Landé factor. Likewise, matrix elements of **μ** and another dipole operator differ by a constant, which is a ratio of reduced matrix-elements. In particular, amplitudes for magnetic neutron diffraction and resonant x-ray diffraction by ferric ions ($^6$S) are proportional to matrix elements of **μ**. This is because higher-order multipoles engaged in diffraction − quadrupoles and octupoles − vanish for a s-state ion. Departure from the high-spin state allow these multipoles to take values different from zero. The crystalline electric field experienced by a ferric ion in magnetic $ScFeO_3$ has no symmetry, and this alone admits the possibility that more than one J-manifold occurs in the magnetic ground-state. In which case, the magnetic moment cannot be calculated from the operator equivalent **μ** = g**J**, and quadrupoles and octupoles are different from zero [11]. A multipole of rank K has projections − K ≤ Q ≤ K. Absence of site symmetry equates to no additional constraint on Q, whereas the triad axis of rotation symmetry experienced by Fe ions in R3c demands Q = 0, ±3, ±6, ... .

## IV. RESONANT X-RAY BRAGG DIFFRACTION

Amplitudes for Bragg diffraction enhanced by an electric dipole resonance, E1-E1, are calculated allowing for quadrupoles forbidden for an s-state ion, $^6S$. It is shown that intensity of a Bragg spot as a function of rotation about the Bragg wavevector (an azimuthal-angle scan) can be used to distinguish between Cc and Cc', in the same manner as Chapon and Lovesey proposed to determine the moment direction from a study of an iridate [3].

States of polarization in Bragg diffraction are displayed in Figure 2, along with a definition of the Bragg angle θ, the nominal setting of the crystal and the Bragg wavevector **τ**(h, k, l). Unit-cell structure factors are expressed in terms of quantities $A^K_Q$ and $B^K_Q$ derived from the electronic structure-factor (A.2).

We consider Bragg spots indexed by (0, 0, l) with l odd that are forbidden in the R3c space group. The reciprocal lattice vector **τ**(0, 0, l) is parallel to the hexagonal c-axis. Evidence for the assumed magnetic motif (G-type antiferromagnetic order) includes the observation of space-group forbidden Bragg spots (0, 0, 1) and (h, k, 1), with h + k = even, that demonstrate magnetic characteristics [1]. The origin of the azimuthal-angle scan (ψ = 0) for the (0, 0, l) reflection places the reciprocal lattice vector $a_m$* normal to the plane of scattering, parallel to σ-polarization. Structure factors for Bragg spots other than (0, 0, l), and other resonant events − E1-M1, E1-E2 and E2-E2 − are readily obtained from general expressions derived by Scagnoli and Lovesey [12, 13].

Intensity of a Bragg spot is proportional to $|F|^2$, where F is a unit-cell structure factor. For the rotated (π'σ) and unrotated (σ'σ) channels of polarization in diffraction by ions occupying sites 4a in Cc,

$F_{\pi'\sigma}(Cc) = - i \sqrt{(1/2)} \cos(\theta) \sin(\psi) A^1_0 + i \sin(\theta) B^1_1$

$- i \sin(\theta) \cos(2\psi) A^2_1 + i \cos(\theta) \sin(\psi) B^2_2,$

$F_{\sigma'\sigma}(Cc) = - i \sin(2\psi) A^2_1,$ (4)

where,

$A^1_0(Cc) = c \langle T^1_\zeta \rangle + s \langle T^1_\xi \rangle$, $B^1_1(Cc) = - \sqrt{(1/2)} [s \langle T^1_\zeta \rangle - c \langle T^1_\xi \rangle]$,

$A^2_1(Cc) = -i [c \langle T^2_{+1} \rangle'' - s \langle T^2_{+2} \rangle'']$, $B^2_2(Cc) = i [s \langle T^2_{+1} \rangle'' + c \langle T^2_{+2} \rangle'']$. (5)

In these and subsequent expressions for $A^K_Q$ and $B^K_Q$ we omit a factor 4, and use c = cos(β) and s = sin(β). As previously mentioned, dipole moments within a J-manifold are proportional to the magnetic moment, e.g, for J = 5/2 appropriate to $^6$S one has $\langle\mu_\zeta\rangle/\langle T^1_\zeta\rangle = -(\pm 45\sqrt{2})$ where the upper sign (lower sign) is for the $L_3$ absorption edge ($L_2$). In which case, observations should tally with $B^1_1(Cc) = 0$, because $B^1_1(Cc)$ is proportional to the magnetic moment parallel to the hexagonal c-axis, together with $A^2_1(Cc) = B^2_2(Cc) = 0$, because quadrupoles are zero for $^6$S. Observations that contradict these results can be understood as a breakdown of the high-spin configuration of the ferric ion. Note that dipoles and quadrupoles contributions to $F_{\pi'\sigma}(Cc)$ are 90º out of phase, which is a manifestation of the fact that dipoles are time-odd (magnetic) and quadrupoles are time-even (non-magnetic).

Corresponding results for Cc′ are,

$F_{\pi'\sigma}(Cc') = \cos(\theta)\cos(\psi) A^1_1$

$- i \sin(\theta) \cos(2\psi) A^2_1 + i \cos(\theta) \sin(\psi) B^2_2,$

$F_{\sigma'\sigma}(Cc') = F_{\sigma'\sigma}(Cc),$  (6)

where,

$A^1_1(Cc') = - i \sqrt{(1/2)} \langle T^1_\eta\rangle,\ A^2_1(Cc') = A^2_1(Cc),\ B^2_2(Cc') = B^2_2(Cc).$  (7)

Time-even contributions to structure factors are the same in Cc and Cc′, as expected. And the dipole contributions differ in a manner that might also be anticipate, for they are shifted in $\psi$ by 90º.

The two space groups have different magneto-electric properties revealed in Bragg diffraction enhanced by a parity-odd event. This type of Bragg diffraction entails Dirac multipoles that are parity-odd and time-odd. As with the electric dipole moment $\langle\mathbf{n}\rangle$, Dirac multipoles may assume non-zero values when ions occupy sites devoid of a centre of inversion symmetry. In such a case the crystalline electric field experienced by an ion can mix orbitals with opposite parities, e.g., 3d and 4p. The predicted strong Fe-O hybridization is a factor in this discussion, because in the vicinity of the cation O-2p states can be represented by occupancy of Fe-4p states [2]. Dirac multipoles have been observed in a variety of materials, including, CuO, $V_2O_3$, and high-Tc cuprates [14-18].

Dirac monopoles that contribute in E1-M1 are allowed for Miller indices $h + l$ odd (even) in Cc (Cc′) [13]. Unit-cell structure factors can be calculated with the information provided in the Appendix in conjunction with published universal expressions for E1-M1 and E1-E2 events [12, 13]. Dirac multipoles, apart from the monopole, also deflect neutrons to which we turn attention in the next section.

## V. NEUTRON BRAGG DIFFRACTION

Neutrons are diffracted by the dipole magnetic moment $\mathbf{\mu} = \langle(\mathbf{L} + 2\mathbf{S})\rangle$, which is the principal reason why neutron Bragg diffraction is the method of choice for the determination of magnetic structures. Dirac multipoles also diffract neutrons, and participating dipoles are already encountered, namely, $\langle\mathbf{n}\rangle$, $\langle\mathbf{S \times n}\rangle$ and $\langle\mathbf{\Omega}\rangle = \langle(\mathbf{L \times n} - \mathbf{n \times L})\rangle$ [19].

The parity-even amplitude can be written $\mathbf{Q}_\perp^{(+)} = [\mathbf{\kappa} \times (\mathbf{Q}^{(+)} \times \mathbf{\kappa})]$ where $\mathbf{\kappa}$ is a unit vector in the direction of the Bragg wavevector, $\mathbf{\tau}(h, k, l)$. For the intermediate operator,

$$\langle \mathbf{Q} \rangle^{(+)} = (1/2) \langle j(k)_0 \rangle \langle (\mathbf{L} + 2\mathbf{S}) \rangle, \tag{8}$$

is an exact result for the particular case of an s-state ion, e.g., $^6$S. In this expression, k is the magnitude of $\tau(h, k, l)$, and $\langle j(k)_0 \rangle$ is a standard radial integral defined such that $\langle j(0)_0 \rangle = 1$ [20]. The expectation value $\langle \mathbf{Q} \rangle^{(+)}$ contains higher-order multipoles for all atomic configurations other than a s-state, and they are accompanied by radial integrals $\langle j(k)_n \rangle$ that vanish at k = 0. Only states with the same parity make contributions to $\langle \mathbf{Q} \rangle^{(+)}$. Multipoles with an even rank, quadrupoles (K = 2) and hexadecapoles (K = 4), measure the mixing of J-manifolds in the atomic state, because they are zero within a J-manifold [11].

The parity-odd amplitude at the level of approximation that includes electronic dipoles (anapoles) is [19],

$$\langle \mathbf{Q} \rangle^{(-)} \approx i \, \boldsymbol{\kappa} \times \langle \mathbf{D} \rangle, \tag{9}$$

where,

$$\langle \mathbf{D} \rangle = (1/2)[\, i(g_1) \langle \mathbf{n} \rangle + 3(h_1) \langle \mathbf{S} \times \mathbf{n} \rangle - (j_0) \langle \boldsymbol{\Omega} \rangle\,]. \tag{10}$$

Dipoles, $\mathbf{n}$, $\mathbf{S} \times \mathbf{n}$ and $\boldsymbol{\Omega}$ are Hermitian. Cartesian components of Hermitian dipoles are purely real, which means that a Cartesian component of $\langle \mathbf{D} \rangle$ is complex when $\langle \mathbf{n} \rangle$ is different from zero. Anapoles $\langle \mathbf{S} \times \mathbf{n} \rangle$ and $\langle \boldsymbol{\Omega} \rangle$ are time-odd, and $i\langle \mathbf{n} \rangle$ has a like property. Radial integrals $(g_1)$, $(h_1)$, $(j_0)$ are significantly different from $\langle j(k)_n \rangle$, as might be anticipated from the fact that they are constructed from radial wavefunctions that belong to states with opposites parities [19].

Neutron polarization analysis can be used to extract the magnetic contribution to the intensity of a Bragg spot with overlapping nuclear and magnetic amplitudes. Primary and secondary polarizations are $\mathbf{P}$ and $\mathbf{P}'$, and a fraction $(1 - \mathbf{P} \cdot \mathbf{P}')/2$ of neutrons participate in events that change (flip) the neutron spin orientation. For a collinear magnetic motif one finds $(1 - \mathbf{P} \cdot \mathbf{P}')/2 \propto \{(1/2)(1 + P^2)|\langle \mathbf{Q}_\perp \rangle|^2 - |\mathbf{P} \cdot \langle \mathbf{Q}_\perp \rangle|^2\}$. A quantity SF = $\{|\langle \mathbf{Q}_\perp \rangle|^2 - |\mathbf{P} \cdot \langle \mathbf{Q}_\perp \rangle|^2\}$ obtained with $P^2 = 1$ is a convenient measure of the strength of spin-flip scattering.

We consider $(h, 0, l)$ Bragg spots in the magnetic diffraction of neutrons; $h$ is even while $l$ may be even or odd. The Bragg wavevector $\tau(h, 0, l)$ is perpendicular to a glide plane. Animadvert that the product of the time-signature and $(-1)^l$ determines whether the electronic structure-factor (A.2) is an even or odd function of the projection Q. Whence, diffraction amplitudes for Cc with $l$ even (odd) and Cc' with $l$ odd (even) are trivially related. Specifically, $\langle \mathbf{Q}_\perp \rangle = [\langle \mathbf{Q}_\perp \rangle^{(+)} + \langle \mathbf{Q}_\perp \rangle^{(-)}]$ for Cc with $l$ even and Cc' with $l$ odd are the same apart from a minus sign. Amplitudes are here listed for Cc with $h = 2n$ and $l$ even or odd.

$l$ even: Reflections are space-group allowed. Results for the parity-even intermediate amplitude are $\langle \mathbf{Q}_\xi \rangle^{(+)} \approx 0$, $\langle \mathbf{Q}_\zeta \rangle^{(+)} \approx 0$ correct to the level of octupoles (K = 3), while

$$\langle \mathbf{Q}_\eta \rangle^{(+)} = \langle \mathbf{Q}_{\perp,\eta} \rangle^{(+)} \approx (-1)^n \,[(3/2)\langle t^1_\eta \rangle + \sqrt{3}\{\kappa_\xi \kappa_\zeta \,[\langle t^2_{+2} \rangle' - \sqrt{(3/2)} \langle t^2_0 \rangle]$$

$$+ (\kappa_\xi^2 - \kappa_\zeta^2) \langle t^2_{+1} \rangle'\}], \tag{11}$$

with $\kappa_\xi \propto h$ and $\kappa_\zeta \propto l$, includes all dipoles and quadrupoles. Parity-even multipoles that arise in neutron scattering, denoted here by $\langle t^K_Q \rangle$, are time-odd (magnetic). (This contrasts with parity-even multipoles in resonant x-ray Bragg diffraction, enhanced by E1-E1 or E2-E2 events, that possess a time signature $\sigma_\theta = (-1)^K$.) The dipole in (11) is a linear combination of radial integrals $\langle j(k)_0\rangle$ and $\langle j(k)_2\rangle$ while $\langle t^2_Q\rangle$ is proportional to $\langle j(k)_2\rangle$. For the corresponding parity-odd amplitude we find $\langle \mathbf{Q}_{\perp,\xi}\rangle^{(-)} \approx 0$, $\langle \mathbf{Q}_{\perp,\zeta}\rangle^{(-)} \approx 0$,

$$\langle \mathbf{Q}_{\perp,\eta}\rangle^{(-)} \approx i\,(-1)^n\,[\kappa_\zeta\,\langle \mathbf{D}_\xi\rangle - \kappa_\xi\,\langle \mathbf{D}_\zeta\rangle + (3/\sqrt{5})\{\kappa_\xi\,\langle H^2_{+2}\rangle'' - \kappa_\zeta\,\langle H^2_{+1}\rangle''\}]. \quad (12)$$

The total diffraction amplitude has one component $\langle \mathbf{Q}_{\perp,\eta}\rangle = [\langle \mathbf{Q}_\eta\rangle^{(+)} + \langle \mathbf{Q}_{\perp,\eta}\rangle^{(-)}]$ with the two contributions taken from (11) and (12). Parity-odd quadrupoles in $\langle \mathbf{Q}_{\perp,\eta}\rangle^{(-)}$ are $\langle H^2_{+1}\rangle'' \propto \langle (S_\eta n_\zeta + S_\zeta n_\eta)\rangle$ and $\langle H^2_{+2}\rangle'' \propto \langle (S_\xi n_\eta + S_\eta n_\xi)\rangle$, and the proportionality includes a radial integral $(h_1)$ that vanishes in the forward direction of scattering. The same radial integral occurs in $\langle \mathbf{D}\rangle$ where it accompanies the anapole $\langle \mathbf{S} \times \mathbf{n}\rangle$.

*l* odd: Reflections are space-group forbidden. Bragg spots $(-2, 0, 1)$ and $(-1, -1, 1)$ have been identified in powder neutron diffraction patterns, with no evidence of a moment parallel to the hexagonal c-axis [1]. The latter observation implies $\langle t^1_\zeta\rangle = \cot(\beta)\,\langle t^1_\xi\rangle$ in Cc, when the extreme case of a s-state ion applies and $\langle \mathbf{t}^1\rangle = [(1/3)\,\langle j(k)_0\rangle\,\langle (\mathbf{L} + 2\mathbf{S})\rangle]$. We go on to find,

$$\langle \mathbf{Q}_\eta\rangle^{(+)} \approx 0,\ \langle \mathbf{Q}_\xi\rangle^{(+)} \approx (-1)^n\,[-(3/2)\,\langle t^1_\xi\rangle +\ \kappa_\zeta\,\sqrt{3}\,\{\kappa_\xi\,\langle t^2_{+2}\rangle'' - \kappa_\zeta\,\langle t^2_{+1}\rangle''\}],$$

$$\langle \mathbf{Q}_\zeta\rangle^{(+)} \approx (-1)^n\,[-(3/2)\,\langle t^1_\zeta\rangle +\ \kappa_\xi\,\sqrt{3}\,\{\kappa_\zeta\,\langle t^2_{+2}\rangle'' - \kappa_\xi\,\langle t^2_{+1}\rangle''\}].$$

$$\langle \mathbf{Q}_{\perp,\eta}\rangle^{(-)} \approx 0,\ \langle \mathbf{Q}_{\perp,\xi}\rangle^{(-)} \approx i\,(-1)^n\,\kappa_\zeta\,[\langle \mathbf{D}_\eta\rangle + (3/\sqrt{5})\,\{\kappa_\xi\,\kappa_\zeta\,(\sqrt{(3/2)}\,\langle H^2_0\rangle - \langle H^2_{+2}\rangle')$$

$$+ (1 - 2\kappa_\xi^2)\,\langle H^2_{+1}\rangle'\}],$$

$$\langle \mathbf{Q}_{\perp,\xi}\rangle^{(-)} \approx i\,(-1)^n\,\kappa_\xi\,[-\langle \mathbf{D}_\eta\rangle + (3/\sqrt{5})\,\{-\kappa_\xi\,\kappa_\zeta\,(\sqrt{(3/2)}\,\langle H^2_0\rangle - \langle H^2_{+2}\rangle')$$

$$+ (1 - 2\kappa_\zeta^2)\,\langle H^2_{+1}\rangle'\}]. \quad (13)$$

The quadrupoles are $\langle H^2_0\rangle \propto \langle (2S_\zeta n_\zeta - S_\eta n_\eta - S_\xi n_\xi)\rangle$, $\langle H^2_{+1}\rangle' \propto \langle (S_\xi n_\zeta + S_\zeta n_\xi)\rangle$ and $\langle H^2_{+2}\rangle' \propto \langle (S_\xi n_\xi - S_\eta n_\eta)\rangle$. The corresponding intensity for diffraction of unpolarized neutrons is,

$$\text{Cc:}\quad |\langle \mathbf{Q}_\perp\rangle|^2 \approx |\langle \mathbf{D}_\eta\rangle|^2 + [\boldsymbol{\kappa} \times \langle \mathbf{Q}\rangle^{(+)}]_\eta\,\{[\boldsymbol{\kappa} \times \langle \mathbf{Q}\rangle^{(+)}]_\eta - 2\langle \mathbf{D}_\eta\rangle''\}, \quad (14)$$

to an accuracy of anapoles in $\langle \mathbf{Q}_\perp\rangle^{(-)}$, and quadrupoles $\langle H^2_Q\rangle$ are ignored. Recall that $\langle \mathbf{D}_\eta\rangle'' = [(1/2)\,(g_1)\,\langle \mathbf{n}_\eta\rangle]$ and the radial integral $(g_1)$ grows in magnitude as the Bragg wavevector diminishes. The real part of $\langle \mathbf{D}_\eta\rangle$ is a linear combination of two anapoles, $\langle \mathbf{S} \times \mathbf{n}\rangle$ and $\langle \boldsymbol{\Omega}\rangle$.

Expression (14) for intensity of $(h, 0, l)$ Bragg spots with *l* odd in diffraction by the motif Cc is to be contrasted with the equivalent intensity created by the motif Cc', namely,

$$\text{Cc':}\quad |\langle \mathbf{Q}_\perp\rangle|^2 \approx |(\kappa_\zeta\,\langle \mathbf{D}_\xi\rangle - \kappa_\xi\,\langle \mathbf{D}_\zeta\rangle)|^2 + (\langle \mathbf{Q}_\eta\rangle^{(+)})^2 - 2\langle \mathbf{Q}_\eta\rangle^{(+)}\,(\kappa_\zeta\,\langle \mathbf{D}_\xi\rangle'' - \kappa_\xi\,\langle \mathbf{D}_\zeta\rangle''), \quad (15)$$

where $\langle \mathbf{Q}_\eta\rangle^{(+)}$ is taken from (11).

## VI. CONCLUSIONS

We report calculations of unit-cell structure factors for Bragg diffraction by magnetic $ScFeO_3$ in the hope that our investigation will stimulate more experimental studies of the interesting multiferroic, which presents both ferroelectric and magnetic polarizations at room temperatures [1]. While the compounds $ScFeO_3$ and $BiFeO_3$ have some properties in common − starting with identical polar chemical structures − they have different magnetic structures

[21] (a G-type magnetic structure found in ScFeO$_3$ [1] can be stabilized in BiFeO$_3$ by an external magnetic field [9] or by chemical doping [10]).

Resonant x-ray diffraction provides a handle on the orientation of the ferric dipole moment in the basal plane of the G-type antiferromagnetic structure. The same experiment offers potential evidence for the departure from a high-spin state. Magnetic neutron diffraction is predicted to offer complementary information. Both types of Bragg diffraction, resonant x-ray and magnetic neutron, probe Dirac multipoles that are allowed because Fe ions do not use sites that are centres of inversion symmetry. A simulation of the electronic structure shows strong Fe-O hybridization [2]. If it exists, the hybridization will surely enhance the magnitude of Dirac multipoles derived from Fe 3d-4p orbitals.

Our calculations are informed by magnetic symmetry. Specifically, magnetic space-groups Cc (Fe magnetic dipoles perpendicular to the glide plane) and Cc' (Fe magnetic dipoles parallel to the glide plane) are derived from the polar chemical structure R3c. The selected monoclinic structures allow the magnetic transition to be continuous [1].

## ACKNOWLEDGMENT

We are grateful to Professor E Balcar (TU, Vienna) for the preparation of Figure 2, and to Dr U Staub for comments after perusing the pre-print. One of us (DDK) acknowledges support from the project TUMOCS. This project has received funding from the European Union's Horizon 2020 research and innovation programme under the Marie Skłodowska-Curie Grant No. 645660.

## APPENDIX

Hexagonal axes are taken to be $a_h$ = a(1, 0, 0), $b_h$ = (a/2)(−1, √3, 0), $c_h$ = c(0, 0, 1), with unit-cell parameters a ≈ 5.1968 Å and c ≈ 13.9363 Å [1]. Integer Miller indices are denoted ($H_o$, $K_o$, $L_o$). The monoclinic unit-cell is derived from the parent structure using a basis {(−2,−1, 0), (0,−1, 0), (2/3, 1/3, 1/3)} and an origin (−1/6,−1/3,1/6). Fe are in sites 4a (1/4, 1/4, 1/2) that possess no symmetry. Unit cell axes are $a_m$ = − (3a/2)(1, 1/√3, 0), $b_m$ = − $b_h$, and $c_m$ = (1/6)(3a, a√3, 2c) while cos(β) = − a[3/{3a$^2$ + c$^2$}]$^{1/2}$ yields β ≈ 122.857°. Principal axes (ξ, η, ζ) use a reciprocal lattice vector $a_m$* ∝ (−c√3, −c, 2a√3), and we choose ξ ∝ $a_m$*, η ∝ $b_m$, ζ ∝ $c_m$.

P1 (#1.1) basis ={(0,1,0),(1,1,0),(1/3,2/3,-1/3)}, origin = (0,0,0), Fe takes site F1 1a(0, 0, 0) and Fe2 1a(1/2, 1/2, 1/2). Unit cell parameters: a ≈ 5.1968 Å , b ≈ 5.1968 Å, c ≈ 5.5301 Å, α ≈ 61.9746°, β ≈ 61.9746° and γ ≈ 60.00°

Electronic properties of Fe ions are encapsulated in multipoles ⟨O$^K_Q$⟩ that are expectation values of spherical tensor-operators of rank K with projections Q that obey the condition − K ≤ Q ≤ K. Examples of dipoles (K = 1) include the magnetic moment **μ** = (**L** + 2**S**) and an orbital anapole **Ω** = (**L x n** − **n x L**), where **S**, **L** and **n** are operators for electron spin, orbital angular momentum and the electric dipole, respectively. Cartesian components of a dipole are O$^1_x$ = (1/√2) (O$^1_{−1}$ − O$^1_{+1}$), O$^1_y$ = (i/√2) (O$^1_{−1}$ + O$^1_{+1}$) and O$^1_z$ = O$^1_0$.

A suitable electronic structure-factor is,

$$\Psi^K_Q = \sum_\mathbf{d} \exp(i\mathbf{d} \cdot \mathbf{k}) \langle O^K_Q \rangle_\mathbf{d}, \qquad (A.1)$$

where $\mathbf{k}$ is the wavevector, and sites labelled $\mathbf{d}$ in a cell are occupied by ferric ions, $(\xi, \eta, \zeta)$. For ions using sites 4a in $Cc'$ the structure factor takes the value,

$$\Psi^K_Q(Cc') = 2(-1)^{n+l}[\langle O^K_Q \rangle + \sigma_\theta \sigma_\pi (-1)^{h+l+K+Q} \langle O^K_{-Q} \rangle]. \qquad (A.2)$$

Miller indices $h = -2H_o - K_o$, $k = -K_o$, $l = (2H_o + K_o + L_o)/3$ obey $h + k = 2n$. The parity of $O^K_Q$ is labelled by $\sigma_\pi = \pm 1$, with $\sigma_\pi = +1$ for an axial operator (e.g., $\boldsymbol{\mu}$) and $\sigma_\pi = -1$ for a polar operator (e.g., $\boldsymbol{\Omega}$). The time-signature $\sigma_\theta = \pm 1$ is absent in $\Psi^K_Q(Cc)$, which is otherwise identical to (A.2). Unit-cell structure factors in the text are expressed in terms of quantities $A^K_Q$ and $B^K_Q$ defined by $\Psi^K_{\pm Q} = A^K_Q \pm B^K_Q$.

.

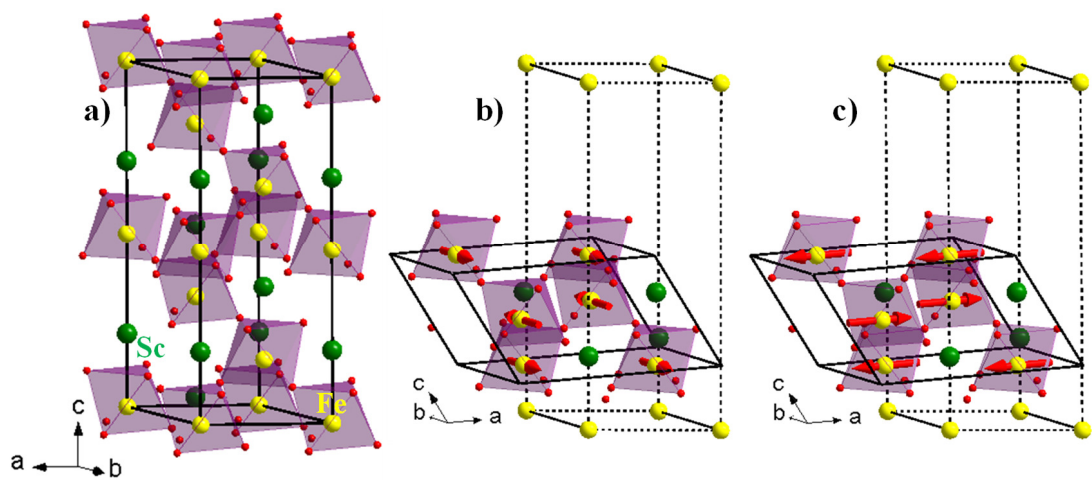

FIG. 1. Unit cell of the paramagnetic crystal structure of ScFeO$_3$ with the R3c1' symmetry (a). Monoclinic unit cells of the magnetic structures with the Cc' (b) and Cc (c) symmetry. In (b) and (c), the monoclinic cell is related to the parent R3c1' structure by {(−2,−1, 0), (0,−1, 0), (2/3, 1/3, 1/3)} with an origin (−1/6,−1/3,1/6).

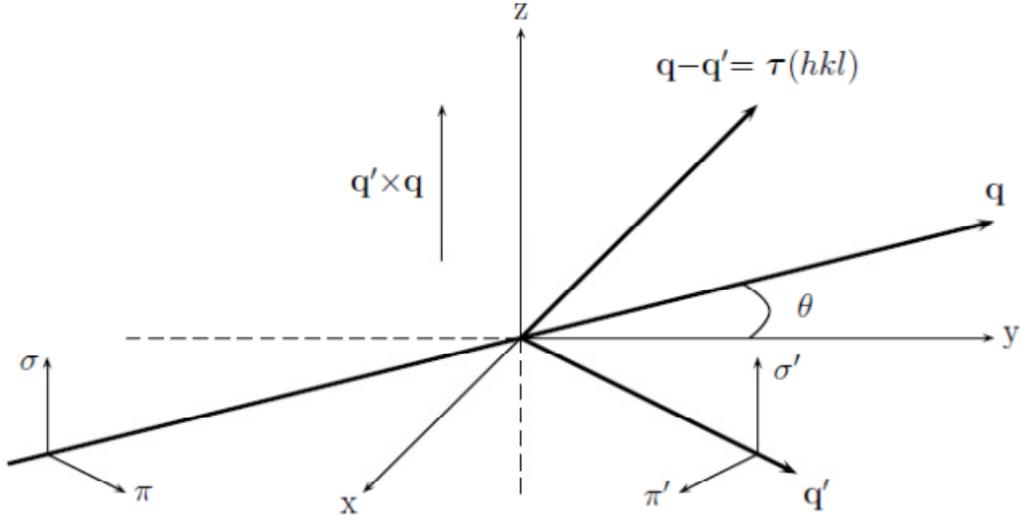

FIG. 2. The diagram illustrates the Cartesian coordinate system (x, y, z) adopted for resonant Bragg diffraction of x-rays and the relation to states of polarization, σ and π, in the primary (unprimed) and secondary (primed) beams. In the nominal setting of the crystal the system (x, y, z) coincides with principal axes (ξ, η, ζ) of a monoclinic cell. At the origin of an azimuthal scan (ψ = 0) a reciprocal-lattice vector $a_m^* \propto (h, 0, 0)$ points along − z. The beam is deflected through and angle 2θ, and **q** and **q'** are primary and secondary wavevectors. Specifically, σ = σ' = (0, 0, 1), **π** = (cos(θ), sin(θ), 0) and **π'** = (cos(θ), −sin(θ), 0).